\documentclass[review]{elsarticle}

\usepackage{lineno,hyperref}
\usepackage[table,xcdraw]{xcolor}
\modulolinenumbers[5]
\usepackage{amsmath}
\journal{Journal of \LaTeX\ Templates}

\begin{document}

\begin{frontmatter}

\title{Mechanical Properties of 3D-Printed Pentadiamond}

\author[1,2]{Levi C. Felix}

\author[3]{Rushikesh S. Ambekar}

\author[4]{Cristiano F. Woellner}
\author[3]{Brijesh Kushwaha}
\author[3]{Varinder Pal}

\author[1,2]{Douglas S. Galvao\corref{corauthor1}}
%\url{galvao@ifi.unicamp.br}

\author[3]{Chandra S. Tiwary\corref{corauthor2}}

\address[1]{Applied Physics Department, 'Gleb Wataghin' Institute of Physics, State University of Campinas, Campinas,SP, 13083-970, Brazil}
\address[2]{Center for Computing in Engineering \& Sciences, State University of Campinas, Campinas, SP, 13083-970, Brazil}
\address[3]{Metallurgical and Materials Engineering, Indian Institute of Technology Kharagpur, Kharagpur, 721302, India}
\address[4]{Physics Department, Federal University of Parana, Curitiba-PR, 81531-980, Brazil}

\cortext[corauthor1]{Corresponding author: galvao@ifi.unicamp.br}
\cortext[corauthor2]{Corresponding author: chandra.tiwary@metal.iitkgp.ac.in}

\begin{abstract}
In this work, We combined fully atomistic molecular dynamics and finite elements simulations with mechanical testings to investigate the mechanical behavior of atomic and 3D-printed models of pentadiamond. Pentadiamond is a recently proposed new carbon allotrope, which is composed of a covalent network of pentagonal rings.  Our results showed that the stress-strain behavior is almost scale-independent. The stress-strain curves of the 3D-printed structures exhibit three characteristic regions. For low-strain values, this first region presents a non-linear behavior close to zero, followed by a well-defined linear behavior. The second regime is a quasi-plastic one and the third one is densification followed by structural failures (fracture). The Young's modulus values decrease with the number of pores. The deformation mechanism is bending-dominated and different from the layer-by-layer deformation mechanism observed for other 3D-printed structures. They exhibit good energy absorption capabilities, with some structures even outperforming kevlar. Interestingly, considering the Ashby chart, 3D-printed pentadiamond lies almost on the ideal stretch and bending-dominated lines, making them promising materials for energy absorption applications.
\end{abstract}

\begin{keyword}
3D Printing \sep Pentadiamond \sep Molecular Dynamics \sep Finite Elements
\end{keyword}

\end{frontmatter}

%\linenumbers

\section{Introduction}

Finding load-bearing, impact-resistant and energy-absorbing materials is of central interest in many areas including aerospace, automotive, civil, sport, packing, and biomedical, among others. In some cases, the desired mechanical response is achieved by manipulating the material composition, such as in the production of steel where iron is mixed with a predefined percentage of carbon. Other possible routes are inspired by nature where the geometry and/or topological determine most of the observed mechanical properties, such as bones \cite{oftadeh_2015}, seashells \cite{tiwary_2015}, ladybug legs \cite{peisker_2013}, shark teeth \cite{frazzetta_1988}, woodpecker \cite{yoon_2011} and kingfisher beaks \cite{lee_2016}. However, structures with controlled geometry are not easily reproduced due to the lack of appropriate synthesis techniques. The recent progress in additive manufacturing (AM) methods have paved the way for effectively fabricating structures with complex geometries \cite{meza_2014,meza_2015,yuan_2019,maskery_2017,maskery_2018,alketan_2019,yang_2018,yu_2019,ambekar_2020}. One AM technique widely used is 3D printing. Some recent investigations \cite{tiwary_2015,zhou_2019,yang_2018biomimetic,lei_2019,song_2019} highlighted the role of geometry of natural materials compared to their 3D-printed counterparts.

Another interesting approach consists of using atomic models as input for 3D printing macroscopic structures. First, a surface is obtained by interpolating all atomic positions. Then, a mesh is generated from the obtained surface that will be subsequently used as input for 3D printing fabrication. This procedure is explained in detail in the Materials and Methods section. Some 3D printed structures were found to possess excellent mechanical response under uniaxial compression and ballistic impact while being extremely lightweight. This opens new perspectives for applications that require strong and lightweight materials in structural applications such as airplanes and spacecraft.

Particularly, schwarzites \cite{sajadi_2018} and tubulanes \cite{sajadi_2019} are two families of carbon-based materials that have shown the effectiveness of the concept of atomic-inspired 3D-printed structures mentioned above. Schwarzites are 3D crystalline materials resembling the shape of the Triply Periodic Minimal Surfaces (TPMS) \cite{mackay_1991}. Tubulanes are also 3D crystalline structures based on crosslinked carbon nanotubes \cite{baughman_1993}. Recent investigations have shown that many features observed in molecular dynamics (MD) simulations of carbon-based Schwarzites and Tubulanes are translated into their corresponding 3D-printed structures \cite{sajadi_2018,sajadi_2019,felix_2020DMRS}. In particular, tubulanes present an incredible ballistic impact performance where a high-velocity projectile is stopped by $5$ mm depth in one of the investigated 3D-printed structures \cite{sajadi_2019}. Another interesting feature is that the deformation mechanisms and high strength behavior are very similar for the atomic and 3D-printed macroscopic schwarzites \cite{sajadi_2018}. These observations suggest a geometry/topology-dominated mechanical behavior that is independent of the length scales.

\begin{figure}[ht]
    \centering
    \includegraphics[scale=0.25]{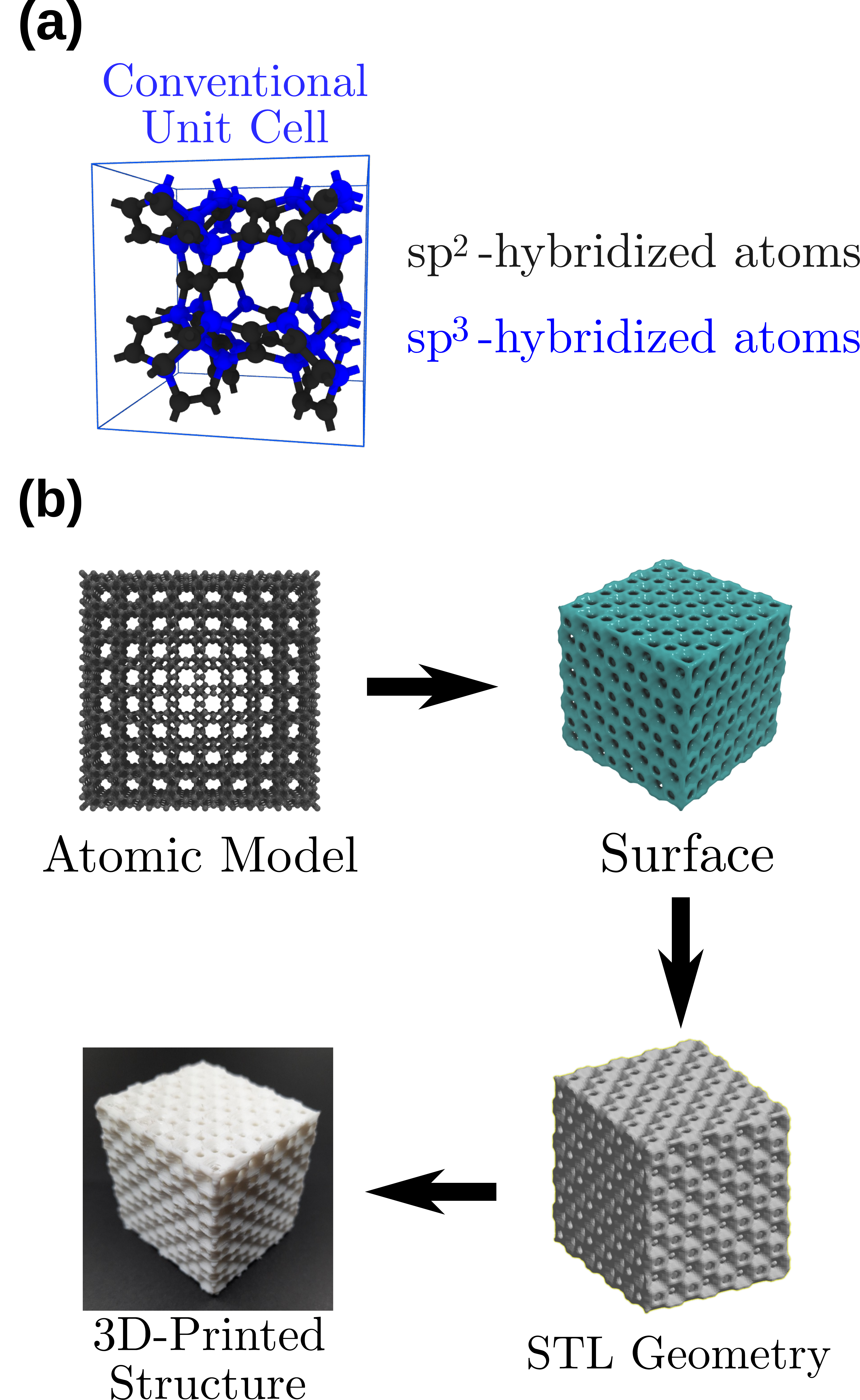}
    \caption{(a) Conventional (cubic) unit cell of pentadiamond where different colors correspond to different atomic hybridizations. (b) Schematic of the procedure used in this work to generate macroscopic 3D-printed structures from atomic models.}
    \label{fig:procedure}
\end{figure}

A feature that schwarzites and tubulanes have in common is their cellular (or porous) structure, which makes them possible good candidates for structural applications \cite{gibson_1997,ashby_2006}. There are mainly two ways in which cellular materials behave under compression: by stretching or bending of the cell walls \cite{ashby_2006}. Stretch-dominated structures have better performance for lightweight structural applications while bending-dominated ones are the best choice for energy absorption applications, such as cushioning, packaging, or to protect against kinetic impact. These definitions are idealizations where a real cellular structure could behave in a manner that is between these two modes of deformation. For an ideally stretch-dominated structure, we have the following scaling behaviors\cite{ashby_2006}
\begin{align}\label{eq:stretch-dominated}
\begin{split}
    \frac{Y}{Y^*} &\approx \frac{1}{3}\frac{\rho}{\rho^*},\\
    \frac{\sigma_Y}{\sigma_Y^*} &\approx \frac{1}{3}\frac{\rho}{\rho^*}.
\end{split}
\end{align}
Here, $\rho$ and $\rho^*$ are the mass density of the cellular structure and the material that it is made of, respectively. The same definition holds for the Young's modulus ($Y$) and the yield strength ($\sigma_Y$). The ideal bending-dominated mode is characterized by the relations\cite{ashby_2006}
\begin{align}\label{eq:bending-dominated}
\begin{split}
    \frac{Y}{Y^*} &\approx \bigg(\frac{\rho}{\rho^*}\bigg)^2,\\
    \frac{\sigma_Y}{\sigma_Y^*} &\approx \bigg(\frac{\rho}{\rho^*}\bigg)^{3/2}.
\end{split}
\end{align}

Another porous carbon material recently proposed was the so-called pentadiamond \cite{fujii_2020,tromer_2020,felix_2021,mortazavi_2021}. It consists of a pentagonal covalent network containing a mixing of sp$^2$ (black) and sp$^3$ (blue) hybridized carbon atoms, as shown in Figure \ref{fig:procedure}(a). Figure \ref{fig:procedure}(b) illustrates the procedure used in this work to obtain the 3D-printed structures. The atomic positions are used to generate a surface that is transformed into a STereoLithography (STL) geometry that is subsequently used to be 3D-printed. Details of the surface generation are explained in the Materials and Methods section.

In this work, we carried out fully atomistic molecular dynamics simulations for the atomic model pentadiamond combined with mechanical testings for macroscale 3D-printed pentadiamond. We investigated the mechanical behavior under compressive loading for the atomic and 3D-printed models. We compared the deformation mechanisms at both scales addressing in particular the role of the topology and number of pores on their mechanical properties.

\begin{figure}[ht]
    \centering
    \includegraphics[scale=0.2]{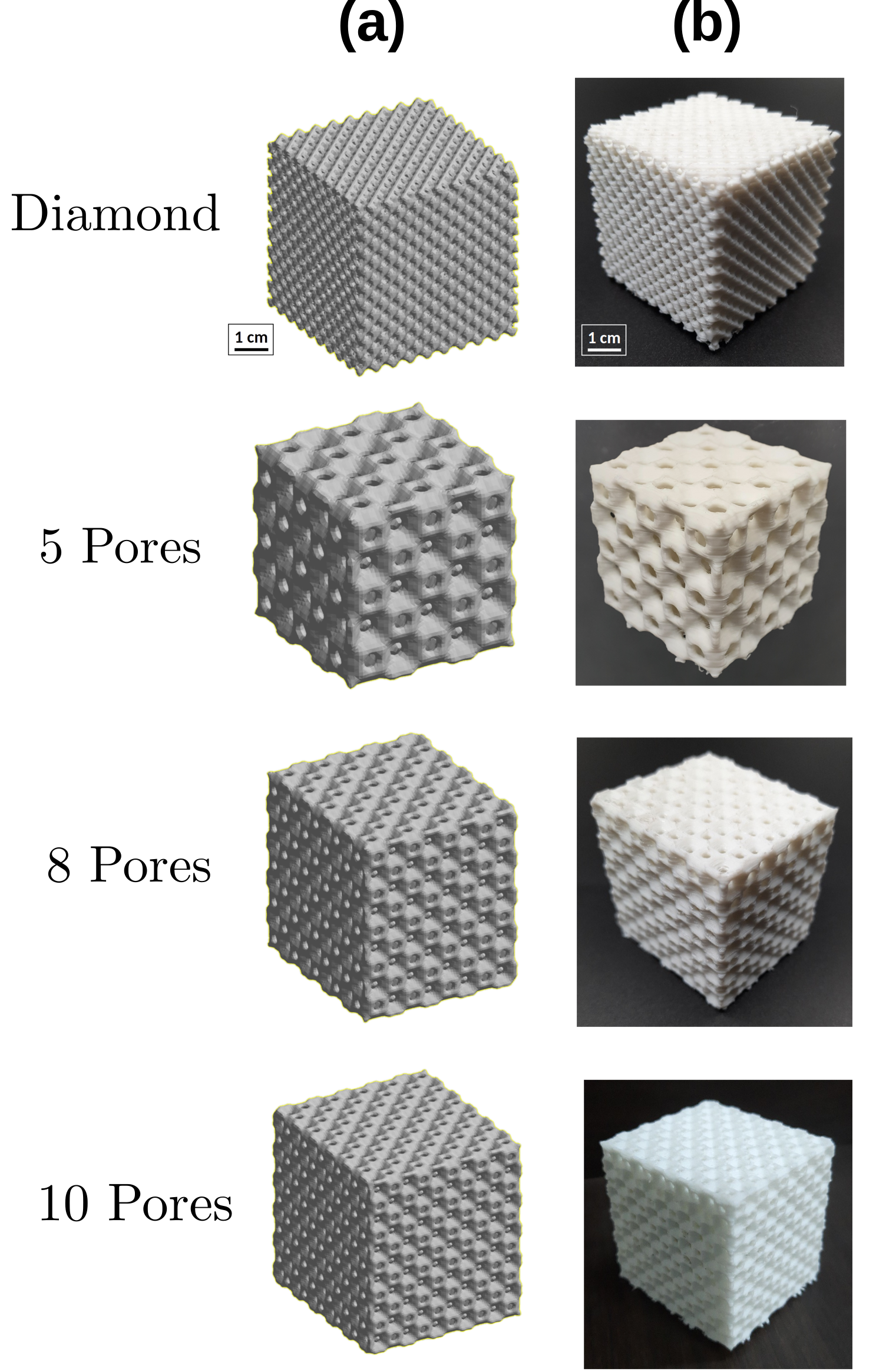}
    \caption{Structures of diamond and pentadiamond studied in this work. (a)  Surfaces generated from the atomic coordinates that were exported to STL files. (b) The corresponding 3D-printed structures fabricated from the geometrical models shown in (a).
}
    \label{fig:structures}
\end{figure}

\section{Materials and Methods}

\textit{Atomistic Simulations:} The atomic models of pentadiamond were generated from the fractional coordinates provided by the paper of Fujii \textit{et al.} \cite{fujii_2020}. In Figure \ref{fig:structures}~(a), it is shown the used pentadiamond supercell sizes of $5\times5\times5$, $8\times8\times8$ and $10\times10\times10$ that we will refer to as 5 pores (5P), 8 pores (8P), and 10 pores (10P), respectively, since these are the number of pores along one of the directions of the Cartesian axes. We considered for the atomic and 3D-printed cases different sizes (number of pores) in order to address size effects on the mechanical behavior. For comparison, we also considered a cubic sample of diamond with 8000 atoms. More details on the studied atomic models, such as the number of atoms, size, and density are listed in Table \ref{tab:structural}. Compressing tests in the atomic structures were performed through Molecular Dynamics (MD) simulations, using the open-source code LAMMPS \cite{plimpton_1995}. The interactions between carbon atoms are described by the reactive potential AIREBO \cite{stuart_2000}. Before compression, a thermalization is carried out for all structures during $100$~ps at a temperature of 1 K. Then, some atoms at the bottom are fixed and a rigid wall (reflecting potential) is moved at a rate of 10$^{-3}$~ps$^{-1}$ to mimic the mechanical compression along the [001] direction ($z$ direction in Figure \ref{fig:structures}~(a)). We used a timestep of $0.1$~fs and a temperature of 1 K in all simulations to reduce thermal fluctuations on the stress-strain (SS) curves. All four atomic structures were simulated in simulation boxes with very large sizes along the orthogonal directions ($x$ and $y$ directions in Figure \ref{fig:structures}~(a)). A vacuum region is used to avoid suppression of lateral deformations during the compression procedures.

\begin{figure}
    \centering
    \includegraphics[scale=0.30]{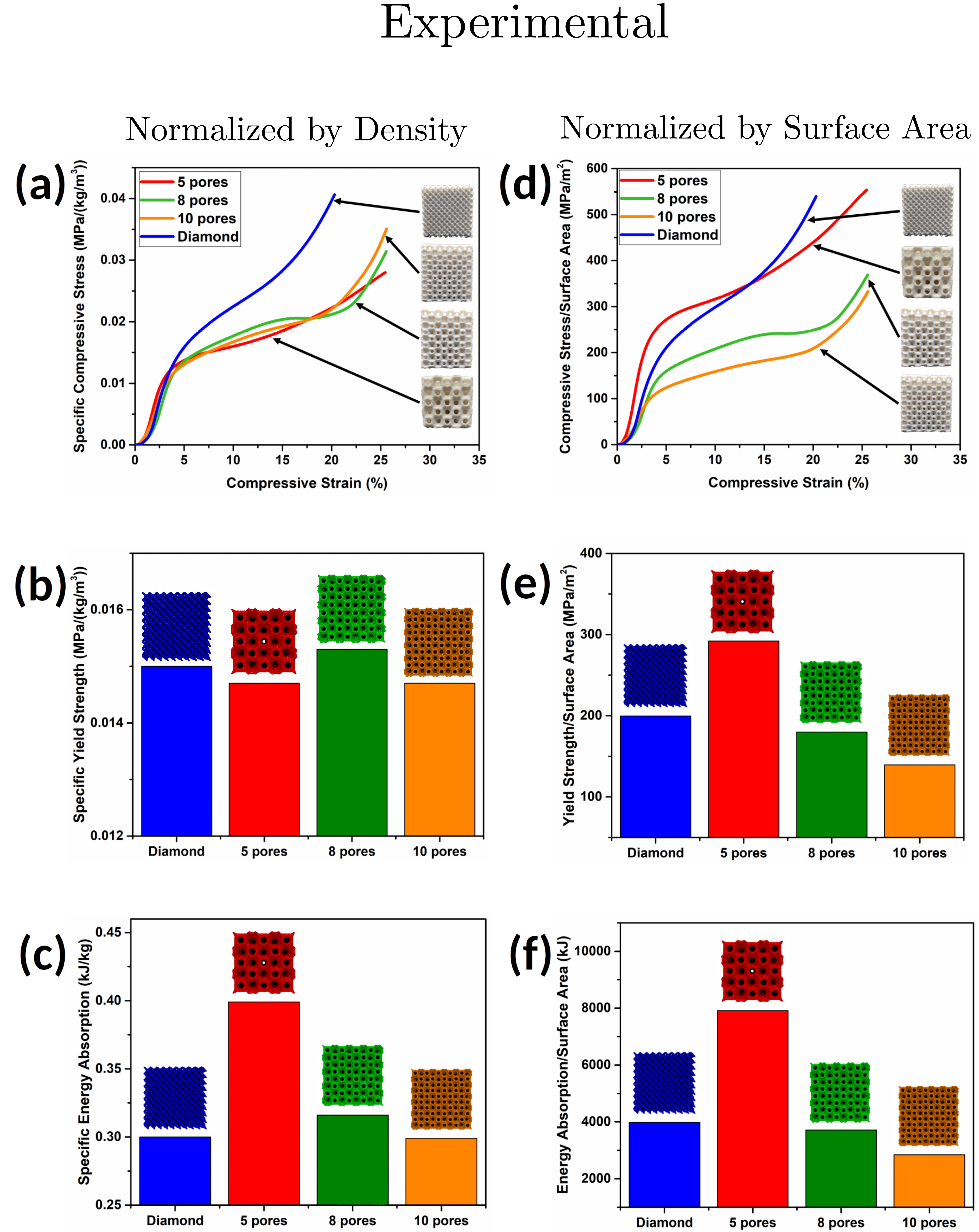}
    \caption{Experimental compression results of 3D-printed diamond and pentadiamond structures normalized by density and surface area. (a) Specific compressive stress/strain curves. (b) Specific yield strength. (c) Specific energy absorption. (d) Compressive stress/surface area versus compressive strain curve. (e) Yield strength/surface area. (f) Energy absorption/surface area.
}
    \label{fig:experimental}
\end{figure}

\textit{Surface generation:} We used the VMD software \cite{humphrey_1996} to obtain a surface representation of the atomic models of both diamond and pentadiamond using the \verb|QuickSurf| representation. For the sake of distinction, we will refer to the 3D-printed structure inspired by diamond as diamond3D. This is based on a  volumetric mapping of the atomic positions into a surface that satisfies a Gaussian density isosurface \cite{krone_2012} of the form

\begin{equation} \label{eq:3}
\rho(\vec{r}, \vec{r}_1, \vec{r}_2, ..., \vec{r}_N) = \sum_{i=1}^{N} e^\frac{-|\vec{r}-\vec{r}_i|^2}{2\alpha^2}
\end{equation}

\noindent
where $\rho$ is the density isovalue, $\alpha$ is the atomic radius scaling factor, and $\vec{r}_i$ are the N atom positions of the atomic model. The user-defined parameters $\rho$ and $\alpha$ give, respectively, how closely packed the surface is and how large is the contribution of atomic volumes on generating the surfaces. In VMD, these contributions are specified by the parameters \verb|Density Isovalue| and \verb|Radius Scale|, which were chosen to be $1.0$ and $0.5$, respectively. The generated surfaces were subsequently exported to STL files, as shown in Figure \ref{fig:structures}~(b), that were used to fabricate the 3D-printed structures shown in Figure \ref{fig:structures}~(c). More details on the 3D-printed structures studied in this work, such as size, density, and porosity are listed in Table \ref{tab:structural}. We have also calculated surface area using 3D-print tool of Blender 2.82 by keeping constant the sizes for all the structures (4 cm$^3$). 

\begin{table}[htb]
\caption{Structural properties of diamond3D and the pentadiamond structures with a different number of pores.}
\label{tab:structural}
\begin{center}
\resizebox{\columnwidth}{!}{%
\begin{tabular}{cccccccccc}
\multicolumn{1}{c}{}          && \multicolumn{3}{c}{MD Simulations} && \multicolumn{4}{c}{3D-Printed Experiments}\\ \cline{3-5}\cline{7-10}
\multicolumn{1}{c}{Structure} && $N$ & $l$ [\AA] & $\rho$ [g/cm$^3$] && $l$ [cm] & $\rho$ [g/cm$^3$] & Porosity [\%] & Surface Area [cm$^2$]\\ \hline 
Diamond3D && 8000 & 34.8 & 3.51 && 4.00 & 0.740 & 40.31 & 557.00 \\ 
\rowcolor[HTML]{C0C0C0}
5 Pores && 2160 & 24.8 & 2.26 && 4.00 & 0.521 & 57.94 & 263.52 \\
\hline 
8 Pores && 7533 & 38.6 & 2.26 && 4.00 & 0.456 & 63.20 & 388.07 \\
\hline
\rowcolor[HTML]{C0C0C0}
10 Pores && 13915 & 47.8 & 2.26 && 4.00 & 0.447 & 63.92 & 471.22 \\
\hline 
\end{tabular}
}
\end{center}
\end{table}

\begin{figure}
    \centering
    \includegraphics[scale=0.30]{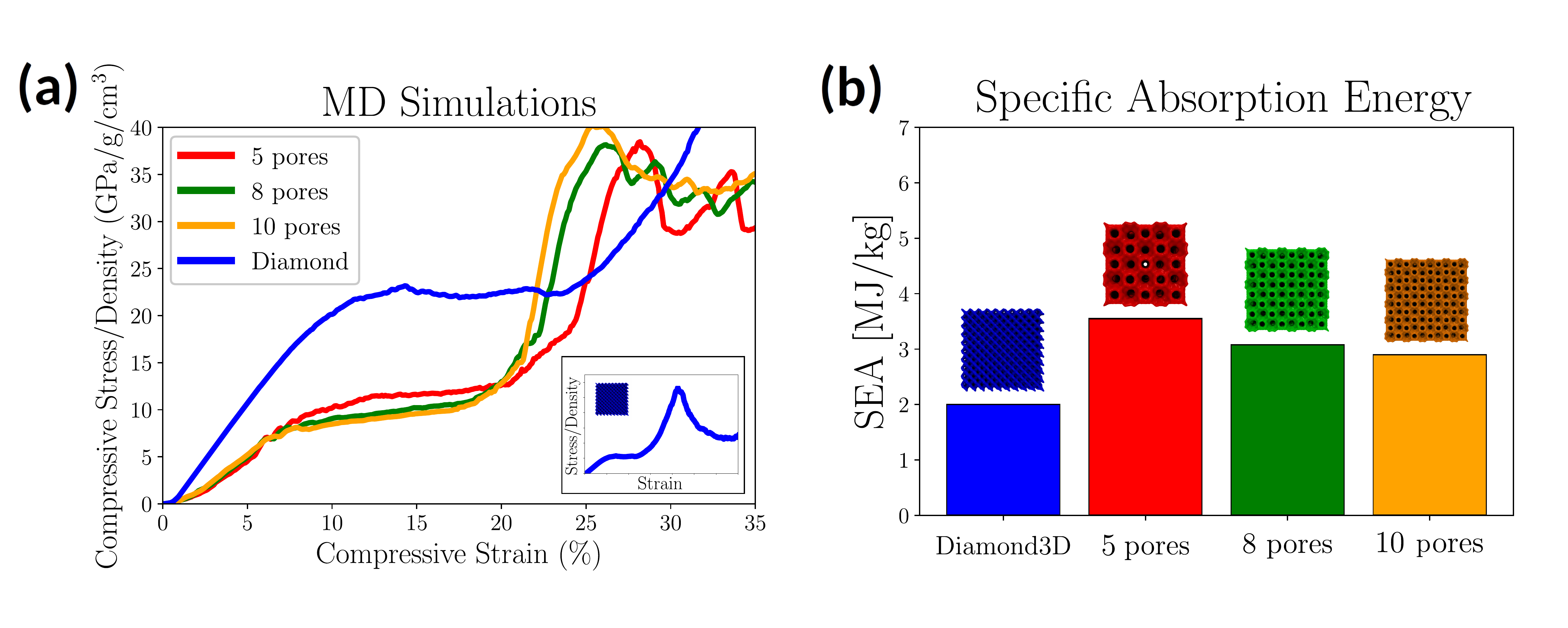}
    \caption{MD simulated compression results of diamond3D and pentadiamond structures (a) Compressive stress/strain curve. (b) Specific energy absorption. In the inset is shown the complete curve for the diamond3D.}
    \label{fig:simulation}
\end{figure}

\textit{3D Printing Fabrication:} The industry-grade PolyLactic Acid (PLA) filament was supplied by Flashforge 3DTechnology Co. Ltd. (Zhejiang, China). The PLA filaments have a low density (1.210-1.430 g/cm$^3$) and standard diameter ($1.75$ mm with tolerance $\pm 0.1$). It melts in the range of $190$ to $220^o$C. All the 3D-printed structures have the same dimensions of $4\times4\times4$ cm. Firstly, the STL files of the structures were processed in Flashprint software and then fed into the 3D printer (Flashforge adventure 3) as a G-code program. The stepper motor causes the movement of extruder assembly along the $x$, $y$, and $z$ directions, as well as the stepper motor also pushes the solid filament into the extruder via a filament guide tube. The extruder temperature was set at $210^o$C and the build platform temperature was set at $50^o$C during the printing process. For ease of removal and good adhesion of structures, paper tape was used. All the printed structures have 100$\%$ fill density and a high resolution (200 $\mu$m single layer thickness). 

\textit{Mechanical Testings:} The compression test of all the structures were carried out along the $z$ direction using the UTM SHIMADZU (AG 5000G). Compression rate (1 mm/min) was maintained constant throughout the test. To analyse the structural deformations with respect to time (increasing strain), we have captured snapshots (30 fps) during the tests.

\textit{Finite Element Analysis:} Finite Element (FE) analysis using ANSYS Workbench 19.2 is performed on the macroscopic models of pentadiamond structures to verify trends on Young's modulus values as a function of porosity. Our structures were modelled by a linear elastic material with Young's modulus of 2.3 GPa, Poisson's ratio of 0.3 and mass density of 1.25 g/cm$^3$, which correspond to PLA. The boundary conditions were chosen to mimic as close as possible the compressive tests. We imposed a displacement of $-0.04$~cm on the top of each structure along the $z$ direction and calculated the distributed von Mises stress and structural deformations.

\textit{Energy Absorption Characteristics}: Since we are studying porous energy-absorbing materials, it is important to calculate the Specific Energy Absorption (SEA), which can be obtained from the area under the strain-strain curve (up to the strain of fracture $\varepsilon=\varepsilon_F$) divided by the structure mass density ($\rho$)
\begin{equation}\label{eq:absorptionW}
    W = \frac{1}{\rho}\int_0^{\varepsilon_F} \sigma(\varepsilon) d\varepsilon.
\end{equation}

\section{Results}

In Figure \ref{fig:experimental} we present the strain/stress (SS) curves and some mechanical/elastic/structural properties of all four structures (diamond3D and pentadiamond with a different number of pores). The SS curves for both 3D-printed structures and some of the corresponding MD results are presented in Figure \ref{fig:experimental} and \ref{fig:simulation}, respectively. The SS curves are normalized by the density and surface area for better comparative study of different density and surface area structures. The SS curves of the 3D-printed structures exhibit three characteristic regions. For low-strain values, this first region presents a non-linear behavior close to zero, which is due to the initial contact between the mechanical press and the structure, which is followed by a well-defined linear behavior. The second region consists of a plateau where the stress increases slowly as the strain increases (quasi-plastic regime). Finally, a steep increase in the stress occurs in the third region, where it is usually called the densification regime. The behavior just described is characteristic of porous/cellular materials \cite{gibson_1997}, such as our 3D-printed pentadiamond samples. Due to their relatively larger pore size, all pentadiamond samples are much softer than the diamond3D structure. 

As illustrated in Figure \ref{fig:experimental}(a), all 3D-printed structures show similar stress-strain behavior except diamond3D structure. Figure \ref{fig:experimental}(b) shows that the specific yield strength of the printed structures, which present similar values, with the 8P presenting the highest yield strength. In Figure \ref{fig:experimental}(c)) we present the specific energy absorption. The 5P exhibits the largest value, which is 33\% larger than the corresponding value for the diamond3D structure.

As can be seen from Table \ref{tab:structural}, the surface area increases from 5P to 10P, but the diamond3D has the largest area. As we normalize the stress-strain curves by the surface area we observe that this has a more significant effect on the curve shapes than the density normalization, as we can see Figure \ref{fig:experimental}(a) and Figure \ref{fig:experimental}(d). It also affects the curve ordering, and 5P exhibits the highest slope. Overall, considering the yield strength (\ref{fig:experimental}(b) and (e)) and energy absorption capability (\ref{fig:experimental}(c) and (f)) for density and surface normalized cases, 5P exhibits the best results. 

In Figure \ref{fig:simulation}(a) we show the stress-strain curves obtained from MD simulations. Similarly to the 3D-print cases (Figure \ref{fig:experimental}) we observed three distinct regimes, but the densification regime results in an abrupt increase in stress values, and the structures ultimately break. The corresponding peak for the diamond3D structure is shown in the inset of Figure \ref{fig:simulation}(a), which occurs at a strain of 42\%, not shown in Figure. All pentadiamond structures present similar stress-strain values where their plateaus are significantly flatter than the one for diamond3D. 

Diamond3D has the lowest energy absorption capability, with $W=2.0$ MJ/kg, where 5P pentadiamond possesses the best performance with $W=3.5$ MJ/kg (Figure \ref{fig:simulation}(b)). Typical values of SEA for automotive crash-worthiness applications are $\sim 0.09$ kJ/kg for epoxy kevlar composites \cite{jacob_2002} and $\sim 0.05$ kJ/kg for TRIP-steel \cite{martin_2010}. Thus, our atomic-inspired structures can be good candidates for energy absorbing applications.

\begin{figure}
    \centering
    \includegraphics[scale=0.1]{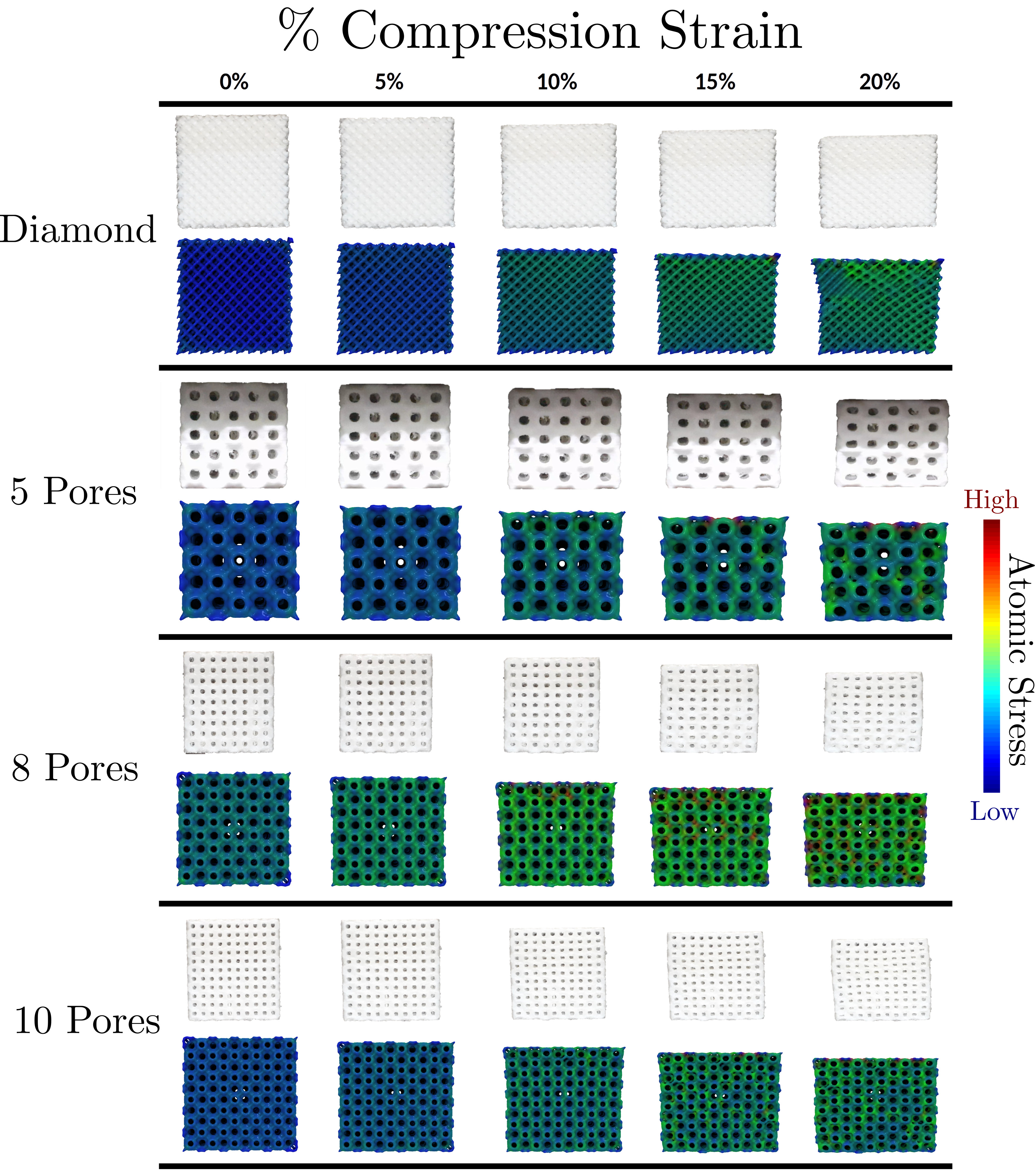}
    \caption{Representative snapshots of the 3D-printed and atomic models at different strain values. For the MD results the von Mises stress values are indicated by the scale bar.}
    \label{fig:snapshots}
\end{figure}

In Figure \ref{fig:snapshots} we present representative snapshots of the 3D-printed and atomic models at different strain values (up to $20\%$). The MD results are presented adopting a surface representation in order to better visualize the atomic stress profiles. From approximately $5\%$ strain (the beginning of the plateau region in Figure \ref{fig:experimental}(a) and (b)) we can see that in Figure \ref{fig:snapshots}, for all structures, all of the pores start to close in a uniform way for both macroscale and atomic structures. This is different from the layer-by-layer deformation mechanism previously observed in the collapse plateau of other atomic-inspired 3D-printed structures \cite{sajadi_2018,sajadi_2019,felix_2020DMRS}. The similarity of the compression stages in Figure \ref{fig:snapshots} show an agreement in the deformation mechanisms for pentadiamond and diamond3D between two different length scales and different materials. These results suggest that the compressive behavior is determined by surface topology since this is the only common aspect between macroscale 3D-printed structures and atomistic ones. From the MD simulations results shown in Figure \ref{fig:snapshots} we can see that the local atomic stress tends to accumulate on the pores since they have large curvatures, which act as stress accumulation regions.

\begin{table}[htb]
\caption{Elastic Properties of investigated structures at $T=1$~K: Young's modulus ($Y$), Poisson's ratio $\nu$, fracture strain $\varepsilon_{F}$ and compressive strengh ($\sigma_S$). Mechanical Properties of finite structures at $T=1$~K: Young's modulus ($Y$), Poisson's ratio $\nu$, fracture $\varepsilon_{F}$ and compressive strengh ($\sigma_S$).}
\label{tab:mechanical}
\begin{center}
\resizebox{\columnwidth}{!}{%
\begin{tabular}{ccccccccccccc}
\multicolumn{1}{l}{}          && \multicolumn{5}{c}{MD Simulations} && \multicolumn{3}{c}{3D-Printed Experiments}\\ \cline{3-7}\cline{9-11}
\multicolumn{1}{c}{Structure} && $Y$ [GPa] & $\sigma_Y$ [GPa] & $\sigma_S$ & $\varepsilon_F$ [\%] & $W$ [MJ/kg] && $Y$ [MPa] & $\sigma_Y$ [MPa] & $W$ [kJ/kg] \\ \hline 
Diamond3D && 858.17 &54.83 & 397.29 & 42.51 & 2.00 && 437.46 & 11.12 & 0.30 \\  
\rowcolor[HTML]{C0C0C0}
5 Pores && 287.62 & 12.49 & 86.95 & 28.19 & 3.55 && 318.43 & 7.68 & 0.40 \\
\hline 
8 Pores && 315.09 & 14.40 & 86.23 & 26.19 & 3.08 && 240.98 & 6.98 & 0.32 \\
\hline
\rowcolor[HTML]{C0C0C0}
10 Pores && 315.40 & 16.17 & 90.84 & 25.38 & 2.90 && 222.93 & 6.57 & 0.30 \\
\hline
\end{tabular}
}
\end{center}
\end{table}

\begin{figure}[ht]
    \centering
    \includegraphics[scale=0.20]{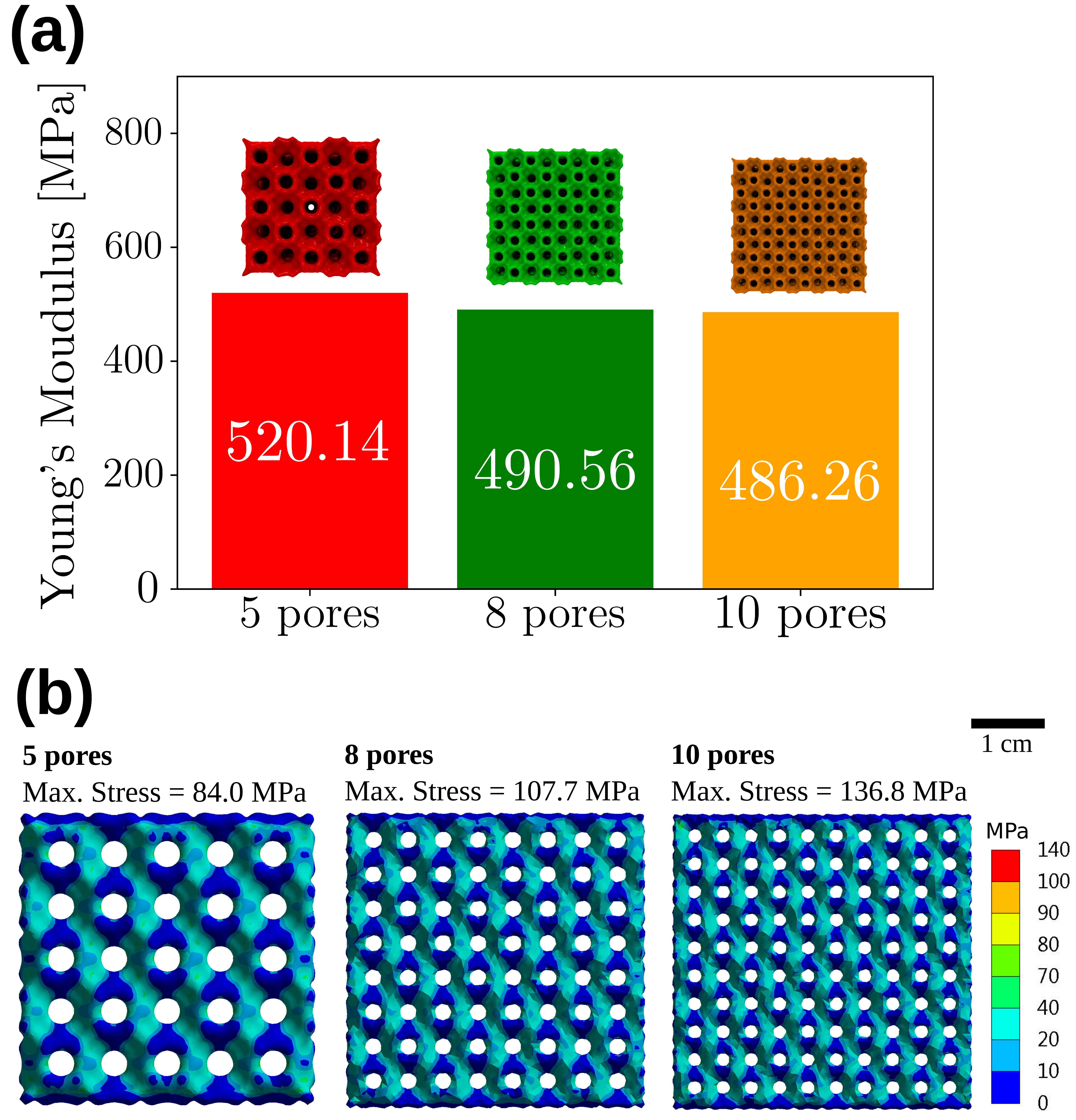}
    \caption{Finite element (FE) analysis results of surface distribution of von Mises stress on (a) 5P, (b) 8P and (c) 10P.}
    \label{fig:fea}
\end{figure}

Young's modulus ($Y$) values are also presented in Table \ref{tab:mechanical}. We would like to stress that a direct comparison between the 3D-printed and MD results for the Young's modulus values is not possible because the 3D-printed and the atomic models have different densities (Young's modulus is density dependent). To better address this scale problem we also carried out a FE analysis the pentadiamond structures within the linear elasticity. In Figure \ref{fig:fea}(a) we present the obtained Young's modules values, which follows the same patterns observed for the 3D-printed structures, i. e., increasing the number of pores decreases the Young's modulus values. In Figure \ref{fig:fea}(b), we present the FE results for the von Mises stress values. The observed values are again consistent with the above discussions for the MD and the mechanical tests, thus validating our conclusions.

\begin{figure}[ht]
    \centering
    \includegraphics[scale=0.35]{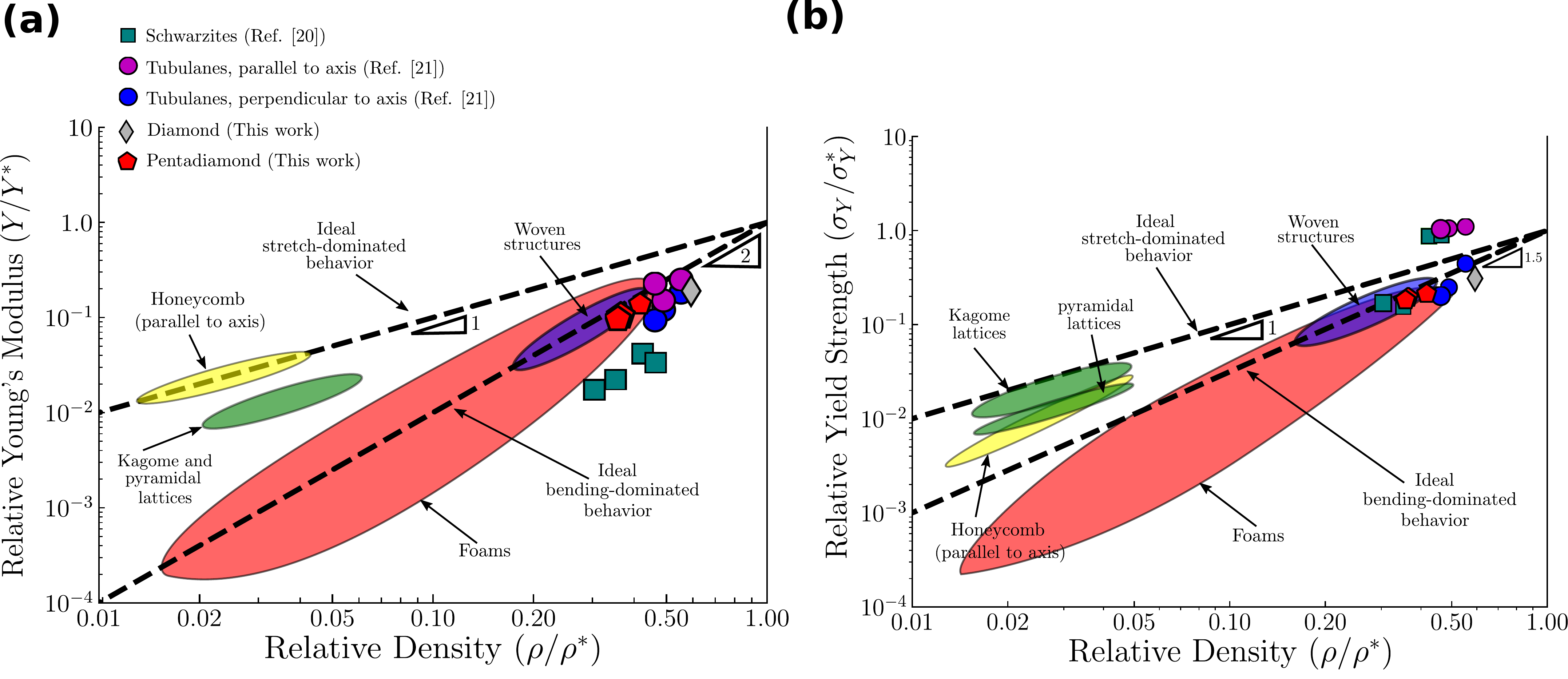}
    \caption{Ashby chart \cite{ashby_2006} for (a) the relative Young's modulus ($Y/Y^*$) and (b) relative yield strength plotted ($\sigma_Y/\sigma_Y^*$) against the relative density ($\rho/\rho^*$) on logarithmic scales for cellular structures with different topologies. Here, $Y^*$, $\sigma_S^*$ and $\rho^*$ are, respectively, Young's modulus, compressive strength, and density of the material that the structures are made of. Ideally, bending-dominated structures lie along a trajectory of slope 1.5; ideally stretch-dominated structures along a line of slope 1.}
    \label{fig:ashby-diagram}
\end{figure}

Finally, in order to contrast the mechanical features of the 3D-printed diamond3D and pentadiamond to other 3D-printed atomic-inspired structures, we created the Ashby charts for Young's modulus and yield strengths, as presented in Figure \ref{fig:ashby-diagram}. From Figure \ref{fig:ashby-diagram}(a), we can see that diamond3D and pentadiamond structures have considerably higher Young's modulus than Schwarzites  \cite{sajadi_2018} while having approximately the same density and yield strength (from Figure \ref{fig:ashby-diagram}(b)). Also, they have similar Young's modulus values as that of tubulanes \cite{sajadi_2019} but with a lower density. Interestingly, the pentadiamond values lie almost in the curve that represents an ideal bending-dominated behavior, which is another indication of their potential for energy absorption applications. \cite{ashby_2006}.

\section{Conclusions}

We combined fully atomistic molecular dynamics and finite elements simulations with mechanical testings to investigate the mechanical response under compressive strain of atomic and 3D-printed models of pentadiamond with a different number of pores. Pentadiamond is a recently proposed new carbon allotrope.  Our results showed that the stress-strain behavior is almost scale-independent and that surface topology plays a dominant role in their mechanical response. The strain-stress curves of the 3D-printed structures exhibit three characteristic regions. For low-strain values, this first region presents a non-linear behavior close to zero, followed by a well-defined linear behavior. The second region consists of a plateau where the stress increases slowly as the strain increases (quasi-plastic regime). Finally, a steep increase in the stress occurs in the third region, where it is usually called the densification regime, after that fracture is observed. The Young's modulus values decrease with the number of pores. Also, the deformation mechanisms are essentially the same across different scales. From approximately $5\%$ strain, for all structures, all of the pores start to close in a uniform way for both macroscale and atomic structures, which is different from the layer by layer deformation mechanism observed for schwarzites \cite{sajadi_2018} and tubulanes \cite{sajadi_2019}.  They exhibit good energy absorption capabilities, with the pentadiamond with five pores even outperforming the kevlar values $0.40$ and $0.09$ kJ/kg, respectively \cite{jacob_2002}. Interestingly, considering the Ashby chart, 3D-printed pentadiamond lies almost on the ideal stretch and bending-dominated lines, making them promising materials for energy absorption applications, such as cushioning, packaging, or protection against kinetic impacts.

%\section*{Data availability statement}

%The data that support the findings of this study are available from the corresponding author upon reasonable request.

\section*{Acknowledgments}

This work was financed in part by the Coordenacão de Aperfeiçoamento de Pessoal de Nível Superior - Brasil (CAPES) - Finance Code 001, CNPq, and FAPESP. We thank the Center for Computing in Engineering and Sciences at Unicamp for financial support through the FAPESP/CEPID Grants \#2013/08293-7 and \#2018/11352-7. We also thank Conselho Nacional de Desenvolvimento Cientifico e Tecnológico (CNPq) for their financial support.

\bibliography{mybibfile}

\end{document}